\documentclass[10pt,conference]{IEEEtran}
\IEEEoverridecommandlockouts

\usepackage{geometry}
\usepackage{amsmath,amssymb,amsfonts}
\usepackage{algorithmic}
\usepackage{graphicx}
\usepackage{textcomp}
\usepackage{xcolor}
\usepackage{booktabs}
\usepackage{breqn}
\usepackage{tikz}
\usepackage{amsthm}
\usepackage{amsmath}
\usepackage[linesnumbered,ruled]{algorithm2e}
\usepackage{comment}
\usepackage{pst-all}
\usepackage{auto-pst-pdf}
\usepackage{nicematrix}
\usepackage{multirow}

\begin{document}

\title{
The Impact of Load Altering Attacks on Distribution Systems with ZIP Loads
\vspace{-6mm}}
\author{\IEEEauthorblockN{Sajjad Maleki\IEEEauthorrefmark{1}\IEEEauthorrefmark{2}, Shijie Pan\IEEEauthorrefmark{4}, E. Veronica Belmega\IEEEauthorrefmark{3}\IEEEauthorrefmark{2}, Charalambos  Konstantinou\IEEEauthorrefmark{4}, and Subhash Lakshminarayana\IEEEauthorrefmark{1} 
}
\IEEEauthorblockA{
\IEEEauthorrefmark{1} School of Engineering, University of Warwick, Coventry, United Kingdom \\
\IEEEauthorrefmark{2}
ETIS UMR 8051, CY Cergy Paris Universit\'e, ENSEA, CNRS, F-95000, Cergy, France\\
\IEEEauthorrefmark{3}Univ. Gustave Eiffel, CNRS, LIGM, F-77454,  Marne-la-Vallée, France\\
\IEEEauthorrefmark{4} CEMSE Division, King Abdullah University of Science and Technology (KAUST)\\
Email: sajjad.maleki@warwick.ac.uk, subhash.lakshminarayana@warwick.ac.uk
}

\vspace{-3em}} 

\maketitle

\begin{abstract}
Load-altering attacks (LAAs) pose a significant threat to power systems with Internet of Things (IoT)-controllable load devices. This research examines the detrimental impact of LAAs on the voltage profile of distribution systems, taking into account the realistic load model with constant impedance Z, constant current I, and constant power P (ZIP). We derive closed-form expressions for computing the voltages of buses following LAA by making approximations to the power flow as well as the load model. We also characterize the minimum number of devices to be manipulated in order to cause voltage safety violations in the system. We conduct extensive simulations using the IEEE-33 bus system to verify the accuracy of the proposed approximations and highlight the difference between the attack impacts while considering constant power and the ZIP load model (which is more representative of real-world loads).
\end{abstract}
\begin{IEEEkeywords}
Cybersecurity, distribution system, load altering attack, voltage profile, ZIP load, IoT-controllable devices.
\end{IEEEkeywords}

\vspace{-3mm}
\section{Introduction}\label{intro}
The increasing popularity and widespread adoption of Internet-of-Things (IoT)-controllable devices (e.g., smart air conditioners, electric vehicle chargers, etc.) have opened up new attack surfaces to target power grids.
In particular, the vulnerability to load-altering attacks (LAAs) that rely on manipulating IoT-controllable appliances has become a significant concern \cite{mohsenian2011distributed, soltan2018blackiot}. To launch LAAs, the attacker does not need to access any classified data about the power system but only needs to target the IoT-controllable loads,  which have much less protection features as compared to supervisory control and data acquisition (SCADA) systems. 

There has been a growing interest in understanding the impact of LAAs on power system operations in recent years. Reference  \cite{soltan2018blackiot} highlighted that large-scale LAAs can lead to significant disruptions in frequency, line failures, and increased operational costs. While the embedded protection schemes can enable power grids to withstand some adverse consequences, LAAs can still lead to controlled load shedding and bulk power partition \cite{huang2019not}. LAAs can also be used to exploit congestion-based vulnerabilities in distribution systems and affect the energy market \cite{khan2021cyber}. Authors of \cite{lakshminarayana2022load} conducted an analysis of the effects of LAAs under high renewable energy penetration conditions and showed that the adverse effects of LAAs are exacerbated due to the low inertia conditions. 

While the works above analyze a one-time manipulation of the system load, reference \cite{amini2016dynamic} analyzed the effect of the so-called dynamic-LAAs (D-LAA) in which the attacker manipulates the system load over a period of time. They showed that such attacks can potentially destabilize the power grid's frequency control loop.  In \cite{lakshminarayana2021analysis}, the authors have provided an analytical framework to study the effects of D-LAAs and find the buses from which the attacker can launch the most effective LAA.

Recent works have also investigated detecting and mitigating LAAs. Reference \cite{youssef2022detection} proposed a time-delay neural network to detect LAAs in smart grids by observing the grid's load profile, whereas \cite{jahangir2023deep} proposed a convolutional neural network approach to detect LAAs using phase angle and frequency measurements of phasor measurement units (PMUs). A novel economic dispatch approach was introduced in \cite{chu2022mitigating} to guarantee the stability of the system under LAA until the attack has been isolated. In \cite{liu2021robust}, the authors introduce an optimal soft open point deployment to mitigate the effects of LAAs on distribution systems by controlling active and reactive power flows in critical normally-open points. 

However, despite the growing literature on LAAs, the majority of the works in this area have focused on power balance in the transmission grid and the effects of LAA on the system frequency. In contrast, the impact of LAAs on the distribution grid has received little attention (with the exception of a few works such as \cite{liu2021robust}). More importantly, none of these works considers the effects of the load models while analyzing the impact of LAAs. It is important to note that real-life loads exhibit voltage-dependent characteristics that must be taken into consideration \cite{van2007voltage}. In this work, we 
address this research gap by studying the impact of LAAs on the {voltage profile} of distribution systems and conduct a comprehensive analysis considering the constant impedance Z, constant current I, and constant power P  (ZIP) load model, which represents the voltage-dependency of loads. 

However, the non-linear power flow models and the voltage dependency pose significant difficulty in the analysis. To address this challenge, we introduce two approximations --  (i) Linearized distribution flow (LinDistFlow) model to calculate the voltage profile of the system which neglects the power losses of lines \cite{baran1989optimal} (ii) ZP approximation for ZIP load model \cite{nazir2020approximate}. The accuracy of the approximations in estimating the true voltages of the system under LAA is verified by conducting extensive simulations using the IEEE-33 bus system. To summarize, the key contributions of this paper are:\\[2pt]
$\bullet$ Taking the voltage dependency of load demand into account by implementing the ZIP load model in our analysis.\\
$\bullet$ Deriving closed-form expressions to calculate the voltage of buses following an LAA and characterizing the minimum number of devices to be compromised in order to cause voltage safety violations in the distribution system. \\
$\bullet$ Analyzing the effect of the location of LAA on the severity of its consequence on the voltage profile of  distribution systems.

The rest of the paper is organized as follows. Section \ref{sec2} presents the implemented introduces the system model. The effect of LAA on the voltage profile of distribution systems is analyzed in Section \ref{analysis}, followed by the introduction of a closed-form approximation in Section \ref{ZPlinear} for computing the voltage profile of the distribution system with ZIP loads. The numerical results and discussions are presented in Section \ref{results}. Lastly, Section \ref{conclusion} concludes the paper.
\section{Preliminaries}\label{sec2}
In this section, we introduce the system model and describe the load models implemented in this work. Lastly, we delve into the discussion of the LAA model.
\subsection{Power System Model}
We use a connected directed graph $G = \{\mathcal{N}, \mathcal{L}\}$ to represent a distribution system, 
where $\mathcal{N} = \{1, 2, \hdots, N\}$ denotes the set of buses and  $\mathcal{L}$ denotes the set of branches. The distribution system graph has a radial structure is hence a tree. Apart from bus $1$ (the root), each bus is called the 'child' of its adjacent bus closer to bus $1$ by one branch; the latter is called the 'parent' bus of the child. Thus, the set of branches is defined as $\mathcal{L} = \{(\pi_i, i)\ | \ \pi_i, \ i \in \mathcal{N}\}$, $\pi_i$ is parent of $i$. In this system, bus $1$ is the generator bus. Furthermore, for simplicity and to clearly illustrate our results, we assume that there is no distributed generation. 
We denote by $\mathcal{D}_k$ the set of buses which forms the unique path connecting bus $1$ to bus $k$, excluding bus $1$ and including bus $k$. The impedance and the power flow of the line $(\pi_i, i) \in \mathcal{L}$ is denoted by $z_{\pi_i, i} = r_{\pi_i, i} + j\ x_{\pi_i, i}$ {($r_{\pi_i, i}$ is the resistance and $x_{\pi_i, i}$ is the reactance of the line)} and $S_{\pi_i, i} = P_{\pi_i, i} +j\ Q_{\pi_i, i}$ {($S_{\pi_i, i}$, $P_{\pi_i, i}$, and $Q_{\pi_i, i}$ stand for apparent, active, and reactive power flows of the line)} respectively. The load of the bus $i \in \mathcal{N}$ is denoted by $S_i^0$ such that
\begin{equation}
    S_i^0 = P_i^0 +jQ_i^0,
    \label{CP}
\end{equation}
where $S_i^0$, $P_i^0$ and $Q_i^0$ represent the apparent, active and reactive power demands respectively. 
\subsection{Power Flow Model}\label{PF}
\subsubsection{Branch flow model}
The branch flow model represents the full AC power flow and the equations describing the steady state of the system are given by (assuming no distributed generation) \cite{baran1989optimal}:
\begin{equation}
    \sum_{k:i\rightarrow k}S_{i,k}  = S_{\pi_i, i} - z_{\pi_i,i}\ |I_{\pi_i, i}|^2 - S_i^0,
    \label{BFM}
\end{equation}
where $V_{\pi_i} - V_i  = z_{\pi_i, i}\ I_{\pi_i, i},$ $S_{\pi_i, i}  = V_{\pi_i}\ I_{\pi_i, i}^*$; while $I_{\pi_i, i}$ is the current flowing through the branch $(\pi_i,i) \in \mathcal{L}$, $V_i$ \textcolor{black}{and $V_{\pi_i}$ are the voltages of the bus $i \in \mathcal{N}$ and of its parent bus respectively,} and the superscript $(\cdot)^*$ denotes the conjugate of a complex number. 
\subsubsection{LinDistflow (LDF)}
LinDistflow (LDF) is a simplified form of the branch flow model in \eqref{BFM} that ignores the branch power losses \cite{baran1989optimal}.
 Under this model, the voltage drop between two consecutive buses is given by
\begin{equation}
    V_k = \sqrt{V_{\pi_k}^2 - 2\ r_{\pi_k, k}\ P_{\pi_k, k} - 2\ x_{\pi_k, k}\ Q_{\pi_k, k}}.
    \label{dv}
    \end{equation}
LDF makes it possible to obtain a linearized model for computing the voltages of buses by substituting $U_k = V_k^2$.
\subsection{Load Models}
\subsubsection{Constant power load model}
The constant power (CP) load model is described by \eqref{CP}, where the load is assumed to be independent of the bus voltages. 
\subsubsection{ZIP load model}
Real-world loads are voltage-dependent and the ZIP load model is used to show this dependency \cite{van2007voltage}. Under this model, the load at node $i \in \mathcal{N}$ as a function of $V_i$ is given by
\begin{equation}
     S^{ZIP}_i(V_i) = P^0_i (\alpha_p + \beta_p V_i + \gamma_p V^2_i) + j Q^0_i (\alpha_q +\beta_q V_i + \gamma_q V^2_i),
     \label{ZIP}
\end{equation}
where $\alpha_k, \ \beta_k$, and $\gamma_k$, $k \in \{p,q\}$ are the coefficients of the ZIP model for constant power, constant current, and constant impedance respectively; \textcolor{black}{and they are obtained experimentally in \cite{bokhari2013experimental}.} Also we have $\alpha_k + \beta_k + \gamma_k = 1.$ 

In this work, we focus on LAAs in which an attacker with access to a cluster of IoT-controllable devices simultaneously turns them on or off, causing an abrupt change in the load demand. We focus on static LAAs, which are one-time manipulations of the demand. If the attack occurs in bus $a\in \mathcal{N}$ and the additional load demand arising from LAA, (i) considering CP load is given by $S^{A}_a = P^{A}_a + jQ^{A}_a$, and (ii) considering ZIP load is given by 
\begin{equation}
     S^{A_{ZIP}}_a = P^{A}_a (\alpha_p + \beta_p V_a + \gamma_p V_a^2) + j Q^{A}_a (\alpha_q + \beta_q V_a + \gamma_q V_a^2).
     \label{ZIPa}
\end{equation}
 
In distribution systems, arguably the most undesirable consequence of LAAs is the alteration of the voltage profile, which will be the main focus of this work. In particular, the objective of this work is to highlight the difference in the impact of LAAs considering realistic ZIP load model as opposed to the CP model used in several prior works.

We note that the impact of a malicious load alteration on the distribution network voltages can be determined by solving the power flow equations described in \eqref{BFM}. Nevertheless, this approach makes it hard to obtain analytical insights about how the effects of various attacks differ. In what follows, we derive analytically tractable expressions to quantify the impact of LAAs using simplified models that nevertheless provide an accurate estimate of the true impact in practice. 
\section{Quantified Analysis of the Effect of LAA on the Voltage Profile}\label{analysis}
The impact of LAAs in the distribution network depends on two key factors (among others) that are the primary focus of this work:
\begin{itemize}
    \item[1.] Dependence on the underlying load model.
    \item[2.] Dependence on the location (i.e., the bus index) where the LAA occurs. 
\end{itemize}
In this section, we will examine how these two factors affect the voltage profile of the distribution system. Our aim is to obtain closed-form expressions to gain an analytical understanding. To end this, we make two simplifying assumptions: (i) We use the LDF model to approximate the distribution system bus voltages following the LAA; and, (ii) We use an approximate model for ZIP loads called ZP {\cite{nazir2020approximate}}. We start by analyzing the impact of LAAs under the CP load model. 

\subsection{Constant Power Loads} \label{LAA-norm}
{This subsection illustrates how to compute the voltages of buses while there is an LAA in the system.} First, using \eqref{dv}, we can write the voltage of any bus $k \in \mathcal{N}$ in terms of the generator bus voltage as 
\begin{equation}
V_k= \sqrt{V_{1}^2 - 2\sum_{i \in \mathcal{D}_k} \left(r_{\pi_i, i}P_{\pi_i, i} + x_{\pi_i, i}Q_{\pi_i, i}\right)}.
 \label{general-1}
\end{equation}
Based on \eqref{general-1}, if there is an LAA in bus $a$, the voltage of bus $k$ ($V_{k}^a$) can be obtained as
\begin{dmath}
V_{k}^a =  
    \sqrt{V_{1}^2 - 2\Delta_k - 2P^{A}_ar_{k,a} - 2Q^{A}_ax_{k,a}},
    \label{fin1}
\end{dmath}
where $\Delta_k = \sum_{i \in \mathcal{D}_k} \left(r_{\pi_i, i}P_{\pi_i, i} + x_{\pi_i, i}Q_{\pi_i, i}\right),$ $r_{k,a} = \sum_{i \in \{\mathcal{D}_a \cap\mathcal{D}_k\}} r_{\pi_i, i},$ and $ x_{k,a} = \sum_{i \in \{\mathcal{D}_a \cap\mathcal{D}_k\}} x_{\pi_i, i}$.
From $\Delta_k$, we notice that, generally, an LAA at the leaf buses results in a higher drop in the voltage profile of the distribution system since the values of $r_{k,a}$ and $x_{k,a}$ will be higher (notice that when $a$ is a leaf bus, the set $\mathcal{D}_a$ has more elements). In other words, an LAA targeting leaf buses has a more severe impact on the system. 

Using \eqref{fin1}, we can obtain an expression to find the number of devices (e.g., air conditioners) the attacker needs to turn on simultaneously in order to cause system voltage safety violations. Let us denote $V_{th}$ as the voltage safety threshold, $P_D$ and $Q_D$ as the active and reactive powers of each device. Then, using \eqref{fin1}, {\color{black} and following straightforward simplifications,} we obtain (assuming  $V_{1}=1$ p.u.) 
\begin{equation}
   P^{A}_a = \frac{U_{th} -1 + 2\Delta_k}{-2(r_{k,a}+ \frac{Q_D}{P_D}x_{k,a})},
    \label{n_devices}
\end{equation}
where $U_{th} = V^2_{th}.$
Using \eqref{n_devices},  $P^{A}_a/P_D$ gives the number of devices that should be simultaneously switched on/off to achieve the attacker's objective.
\vspace{-1mm}
\subsection{ZIP Loads}
In this subsection, we conduct a similar analysis considering the ZIP load model. The key difference, in this case, is that as the voltage of the system drops due to the LAA, the loads will consume less power \cite{bokhari2013experimental}. As a result, we expect to obtain a better voltage profile (and a lower voltage drop) under the ZIP model.
Using \eqref{fin1} and \eqref{ZIPa}, we obtain the following voltage in bus $k \in \mathcal{N}$ when the LAA takes place in bus $a$
\begin{equation}
     V_{k}^a= \sqrt{V_{1}^2 - 2\Delta^{ZIP}_k - 2P^{A_{ZIP}}_ar_{k,a} - 2Q^{A_{ZIP}}_ax_{k,a}},
    \label{ZIP-gen}
\end{equation}
where
\begin{equation}
    \Delta^{ZIP}_k = \sum_{i \in \mathcal{D}_k} \left(r_{\pi_i, i}P_{\pi_i, i}^{ZIP} + x_{\pi_i, i}Q_{\pi_i, i}^{ZIP}\right),
    \label{NL}
    \vspace{-4mm}
\end{equation}
       \begin{align*}
       \vspace{-0.1in}
     P_{\pi_i, i}^{ZIP} &=  P_{\pi_i, i}(\alpha_{p} + \beta_{p} V_i + \gamma_{p} V_i^2),\\
    Q_{\pi_i, i}^{ZIP} &=  Q_{\pi_i, i}(\alpha_{q} + \beta_{q} V_i + \gamma_{q} V_i^2),\\
   P^{A_{ZIP}}_a &= P^{A}_a(\alpha_{p_a} + \beta_{p_a} V_a + \gamma_{p_a} V_a^2),\\
  Q^{A_{ZIP}}_a &= Q^{A}_a(\alpha_{q_a} + \beta_{q_a} V_a + \gamma_{q_a} V_a^2).
       \end{align*}
We note that in this case, the bus voltages cannot be obtained in closed form because the above load values depend quadratically on the voltages. Thus, we use an iterative algorithm in which to calculate the voltages. \textcolor{black}{This iterative method is based on the backward-forward sweep (BFS) method \cite{shirmohammadi1988compensation} with an additional iteration feature, which updates the loads values in each step based on the previous step calculated voltage. The full description of the algorithm is omitted here because of the lack of space.} We note once again that the iterative algorithm despite its effectiveness does not yield any analytical insights (which is the main objective of this paper). We address this issue in the following section by introducing one more approximation. 

\section{Closed-form Approximation of Voltages Under the ZIP Model} \label{ZPlinear}

In this section, we propose an alternative solution for computing the bus voltages with the ZIP model in one shot or closed form (instead of the iterative procedure) by introducing an additional approximation.

\subsection{No Attack}\label{one_no_attack}
To this aim we exploit the approximate ZP model for the ZIP model introduced in \cite{nazir2020approximate}. This eliminates the $\beta V_i$ term of the load model in \eqref{ZIP} and the resulting load is
\begin{equation}
    S^{ZP}_i(V_i) = P^{0}_i (\alpha'_p  + \gamma'_p V_i^2) + j Q^{0}_i (\alpha'_q + \gamma'_q V^2_i),
     \label{ZP}
\end{equation}
where
$\alpha'_p = \alpha_p + \frac{\beta_p}{2}$
,$\alpha'_q = \alpha_q + \frac{\beta_q}{2}$
,$\gamma'_p = \gamma_p + \frac{\beta_p}{2}$
, and $\alpha'_q = \gamma_q + \frac{\beta_q}{2}$. {Implementing the ZP model in \eqref{general-1} we obtain}
{
\begin{equation}
   V_k= \sqrt{V_{1}^2 - 2\sum_{i \in \mathcal{D}_k} \left(r_{\pi_i, i}P^{ZP}_{\pi_i, i} + x_{\pi_i, i}Q^{ZP}_{\pi_i, i}\right)}, 
\end{equation}
where $ P_{\pi_i, i}^{ZP} =  P_{\pi_i, i}(\alpha'_{p} + \gamma'_{p} V_i^2),$ $ Q_{\pi_i, i}^{ZP} =  Q_{\pi_i, i}(\alpha'_{q} + \gamma'_{q} V_i^2).$ Then, a variable change $(U_k=V_k^2)$ results in a set of linear equations which can be depicted as matrix form as follows
\begin{equation}
    \textbf{U}_{(N-1)\times 1} = \Omega_{(N-1)\times N}\begin{bmatrix}
        1\\
        \textbf{U}
    \end{bmatrix}_{N \times 1},
    \label{lin-ZP}
\end{equation}
where $\textbf{U}$ is the vector of squares of voltages, and $\Omega_{(N-1)\times N}$ is the matrix of entries: 
\begin{equation}
\omega_{i, 1} = 1- \sum_{m \in \mathcal{D}_i} (2r_{\pi_m, m} P_{\pi_m, m}^0 \alpha'_{p} +2x_{\pi_m, m} Q_{\pi_m,m}^0 \alpha'_{q}),
\label{w}
\end{equation}
\begin{equation}
\omega_{i, k} = 
\begin{cases}
\sum_{c=2}^i -2 r_{\pi_c, c} P_k^0 \gamma'_{p} -2 x_{\pi_c, c} Q_k^0 \gamma'_{q}, \ \ \text{if} \hspace{0.2cm} i \in \mathcal{D}_{k}\\
\omega_{\pi_i, k},  \ \  \text{otherwise},
\end{cases}
\label{w'}
\end{equation}
where $P_{\pi_m, m}^0$ and $Q_{\pi_m, m}^0$ are active and reactive powers flowing in the branch $(\pi_m,m)\in \mathcal{L}$ while $V_{m}= 1$ p.u., also $2\leq i \leq N$ and $2\leq k \leq N$. 
To solve the system of linear equations in \eqref{lin-ZP}, we can re-write it as follows:
\begin{equation}
    (\textbf{I}_{(N-1) \times (N-1)} - \Omega'_{(N-1) \times (N-1)}) \ \textbf{U}_{(N-1)\times 1} = \Omega''_{(N-1) \times 1},
    \label{ZP-close}
\end{equation}
where $\Omega''_{(N-1)\times 1} = [\omega_{2,k}]$, $\Omega'_{(N-1) \times (N-1)} = [\omega_{i,k}]$ for $i = \{3,4,...,N\}$, and $k \in \mathcal{N}$.
To sum up, our closed-form or one-shot approximation of the bus voltages is:\\ $\textbf{U}_{(N-1)\times 1} = (\textbf{I}_{(N-1) \times (N-1)} - \Omega'_{(N-1) \times (N-1)})^{-1}\ \Omega''_{(N-1) \times 1}$, \\ assuming that $\mathrm{det}(\textbf{I}_{(N-1) \times (N-1)]} - \Omega'_{(N-1) \times (N-1)}) \neq 0$, which seems to be always the case in our setting given the relatively small values of $\omega_{i,j} \ll 1$ obtained numerically.
\subsection{Under LAA}\label{one_LAA}
Introducing LAA into the distribution system will lead to changes in the coefficients of the proposed closed-form approximation. Here, we can also re-write the voltages in equation \eqref{fin1} as: $\textbf{U}^{A}_{(N-1)\times 1} = \Omega^{A}_{(N-1)\times N}\begin{bmatrix}
        1\\
        \textbf{U}^{A}
\end{bmatrix}_{N \times 1}$.
The new coefficients of the matrix $\Omega^A$ are denoted by $\omega_{i,k}^A$, which comprises two parts: $\omega_{i,k}$ representing the calculated coefficients for normal operation circumstances, and $\omega_{i,k}^a$ representing the additional part resulting from LAA. Considering an LAA in bus $a$, the additional load resulting from the LAA is given by $S_a^{A_{ZP}} = P_a^{A_{0}} (\alpha'_{p}  + \gamma'_{p} U_a) + jQ_a^{A_{0}} (\alpha'_{q}  + \gamma'_{q} U_a)$.

Then, using equation \eqref{fin1} we obtain the following matrix coefficients for $i\geq 2$ and $k\geq 2$:
\begin{equation}
    \omega_{i,1}^a = \sum_{c \in\{\mathcal{D}_i\cap\mathcal{D}_a\}} -2P_a^{A_{0}} \alpha'_{p} r_{\pi_c, c} - 2Q_a^{A_{0}} \alpha'_{q} x_{\pi_c, c} , \label{omega_a_eq}
\end{equation}
\begin{equation}
\omega_{i,k}^a =
\begin{cases}
   -2r_{\pi_i,i}P_a^{A_{0}}\gamma'_{p} -2x_{\pi_i,i}Q_a^{A_{0}}\gamma'_{q},  \ \ \text{if} \hspace{0.2cm}i\in \mathcal{D}_{a}\\
    \omega_{\pi_i,k}^A, \ \  \text{otherwise}.
\end{cases}
\end{equation}
Finally, we obtain $\omega_{i,k}^A = \omega_{i,k} + \omega_{i,k}^a$, for $i\geq 2$ and $k\geq 1$. $\Omega^{A}_{(N-1)\times N} =[\Omega^{A''}_{(N-1)\times 1} \ \Omega^{A'}_{(N-1)\times (N-1)} ]$ is formed by the new coefficients and the new bus voltages can be derived as
\begin{equation}
    \textbf{U}^{A}_{(N-1)\times 1} =(\textbf{I} - \Omega^{A'})^{-1} \Omega^{A''}.
    \label{ZB-A}
\end{equation}
{Solving \eqref{ZB-A} enables calculating the voltages of buses by solving a system of linear equations for a system with LAA.}
\subsection{Number of Devices Required for LAA} \label{C_attacks}
In this subsection, we wish to determine the minimum number of targeted devices in the LAA at leaf buses that are required to cause a voltage drop below the acceptable voltage nadir ($V_{th}$). For this, we utilize the {\eqref{ZB-A}} in Subsection \ref{one_LAA}. In these equations, the voltage value at the leaf bus under attack is known to be $V_{th}$, while the values of $P^{A_0}_a$ and $Q^{A_0}_a$ are unknown.  Consequently, we have $Q^{A_0}_a=\frac{Q_{D}}{P_{D}}\ P^{A_0}_a$. 

Since the voltage value at the leaf bus under attack is known, the linear system allowing us to compute the other bus voltages and $ P^{A_0}_a$ will change as follows. The new coefficients of the $\Omega^{d}$ matrix are: 
\begin{align}
    \omega_{i,1}^{D} & = 
    \begin{cases}
       \omega_{i,1} + V_{th}^2\omega_{i, a}, \ \ \text{if}\ i \neq a,\\
        \omega_{i,1} + V_{th}^2(\omega_{i, a}- 1), \ \ \text{if}\ i = a,    
    \end{cases} \\[2pt]
\omega_{i,k}^{D} & = 
\begin{cases}
    \sum_{c\in \mathcal{D}_i} -2r_{\pi_c,c}\alpha'_{p}-2 \frac{Q_{D}}{P_{
    D}}x_{\pi_c,c}\alpha'_{q}, \ \ \\
    \quad \quad \quad \ \text{if}\  k=a, i\in\mathcal{D}_a,\\
    \omega_{\pi_i,k}^{D}, \ \ \text{if} \ k=a, i\notin\mathcal{D}_a,\\
    \omega_{i,k}, \ \ \text{otherwise}.
    \end{cases}
\end{align}

We define $\textbf{X}$ as a vector with the same dimension as $\textbf{U}$. All of the elements of this vector are the same as $\textbf{U}$ except for one, where $\textbf{X}$ contains $P_a^{A_0}$ instead of $U_a$ (since we already know $U_a = V_{th}^2$).
We can then obtain the matrices $\Omega^{d} = [\Omega^{d''} \Omega^{d'}]$. At last, by solving the linear system of equations
\begin{equation}
   (\textbf{I}' - \Omega^{d'})\textbf{X} = \Omega^{d''},
   \label{n_acs}
\end{equation}
where $\textbf{I}'$ is an identity matrix but the $a^{th}$ element of the main diagonal is zero. we obtain the bus voltages as well as the required $P^{A0}_a$. Then, $Q^{A0}_a$ can be calculated based on this.
\section{Numerical Results and Discussion} \label{results}
All the simulations have been carried out in MATPOWER using the IEEE-33 bus system. For simulations, $P^0_i$ and $Q^0_i$ are $50\%$ of default values in MATPOWER case files. The ZIP load coefficients of all the buses which are set to the residential loads-type F \cite{bokhari2013experimental}. \textcolor{black}{To integrate the LAAs, we employ \eqref{omega_a_eq} - \eqref{ZB-A} using $n_{atk}P^0$ and $n_{atk}Q^0$ as $P^{A_0}_a$ and $Q^{A_0}_a$ respectively, in which $n_{atk}$ is the number of attacked devices. Note that each device has unique $P^0$, $Q^0$ and ZIP coefficients and according to the type of attacked devices, they all are altered based on \cite{bokhari2013experimental}.} Further, we set $V_{th} =0.95$ p.u.
In what follows, we present results on the accuracy of approximations introduced in Sections \ref{PF} and \ref{one_no_attack} and find the critical load to be manipulated under the LAA to cause safety violations in the bus voltages.
\subsection{Accuracy of the Approximations} \label{results_accuracy}
 First, we examine the accuracy of the approximations in computing the bus voltages considering the ZIP load model and ignoring the LAAs. 
Note that the BFS method provides the true voltages since it considers the full AC power flow model in \eqref{BFM}. Considering $V^{BFS}_i$ as the true value for the voltage, the maximum error, {\color{black}$e_i = \frac{|V^{BFS}_i - V_i|}{V^{BFS}_i} \times 100,$} (over all the buses) for the voltages computed using \eqref{ZP-close}, and represented in \textcolor{black}{Fig.} \ref{zip-nozip} was $1.07\%$ {\color{black}(for a base voltage of 0.97 p.u.)}, which is a relatively small value and justifies the use of the approximations.

\subsection{Impact of LAAs} \label{effects}
Next, we evaluate the impact of the location of LAA and the load model under consideration in three scenarios -- (i) no LAA, (ii) LAA at bus~3, and (iii) LAA at bus~18. Based on the configuration of the test system, bus 18 is a leaf bus. For both scenarios with LAA, the victim appliances are chosen to be $800$ air conditioners. To evaluate the attack impact, we compute the voltages using the iterative method.

\textcolor{black}{Fig.} \ref{zip-nozip} illustrates that launching an LAA at a leaf bus has a more severe impact on the voltage profile. However, when we consider the ZIP model, the negative impact of the LAA is reduced while two factors contribute to the distinct voltage profiles observed in the system. The first factor is the location of the LAA, in which case we observe an effect similar to the case of CP loads (i.e., an attack at the leaf bus has a more severe impact). On the other hand, a second factor includes the dependency of the load on the nodal voltages.
\begin{figure}
    \centering
    \vspace{-4mm}
    \includegraphics[width=0.45\textwidth]{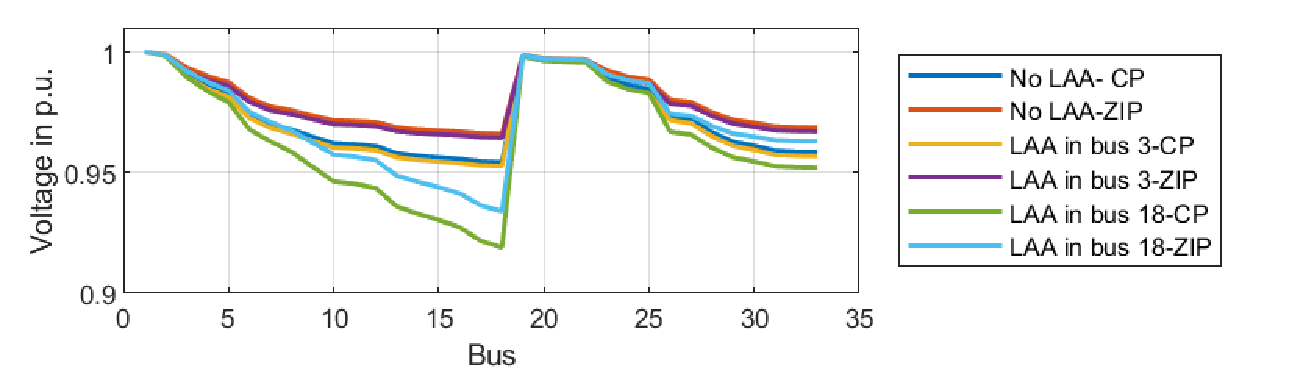}
    \vspace{-4mm}
    \caption{Voltage profile of the IEEE 33-bus system with either CP or ZIP loads under an LAA manipulating $800$ air conditioners.}
    \label{zip-nozip}
    \vspace{-4mm}
\end{figure} 

Table \ref{LAA_S} represents the additional net load demand resulting from LAA in two different attack scenarios. We note that the net load demand considering the voltage dependency is lower when the attack occurs at the leaf buses. Thus we can conclude that the voltage dependency of loads will alleviate the effect of the first factor to some extent.
\begin{table}
    \centering
    \caption{Additional load demand caused by LAAs with ZIP loads.}
    \vspace{-2mm}
    \begin{tabular}{|c|c|c|} \hline
       LAA Bus  &Additional P (kW)& Additional Q(kVAR)\\ \hline
        3 & 395.53 & 97.58 \\ \hline
        18 & 386.09 & 80.20\\  \hline
         \end{tabular}
\vspace{-6mm}
\label{LAA_S}
\end{table}
\subsection{Critical Attacks} \label{critical_attacks}
Finally, we compute the least number of appliances to be compromised to cause system voltage safety violation. Equation \eqref{n_devices} determines the critical device count for the system with CP loads, while \eqref{n_acs} calculates it for the system with ZIP loads. The results are summarized in Table \ref{allbuses} where we inject LAAs at different leaf buses in the system (one at a time). We observe that launching an LAA at bus~$18$ requires the least number of compromised devices. This also confirms the results observed in \textcolor{black}{Fig.} \ref{zip-nozip}, where we similarly observe that launching an LAA at bus $18$ is most detrimental to the system. 

In \textcolor{black}{Fig.} \ref{438}, the voltage profile (computed using the model in Subsection \ref{one_LAA}) of the system is depicted in the presence of an LAA at bus $18$. According to this figure, based on the closed-form approximation, launching LAA on 282 air conditioners, 163 resistive heaters, or 151 copiers in bus $18$ is sufficient to cause voltage constraint violation. 
\begin{figure}
     \centering
     \vspace{-3mm}
     \includegraphics[width=0.45\textwidth]{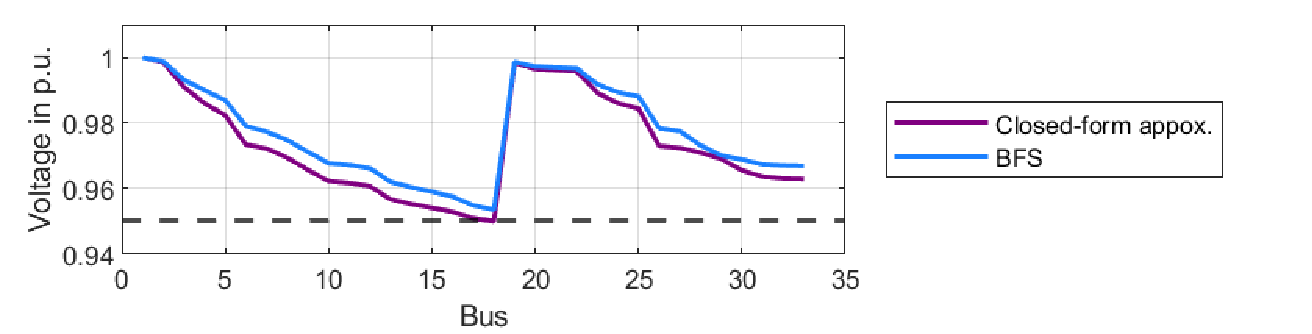}
         \vspace{-4mm}
     \caption{Voltage profile of the system with 282 air conditioners under LAA in bus 18 obtained by proposed closed-form approximation and BFS.}
         \vspace{-4mm}
     \label{438}
\end{figure}

\begin{table}[]
    \centering
    \vspace{-4mm}
    \caption{Required number of devices to be manipulated in different leaf buses to cause a voltage drop below $0.95$ p.u.}
    \vspace{-2mm}
    \begin{tabular}{|l|l|l|l|l|l|l|} \hline
       \multirow{2}{*}{Bus}&\multicolumn{2}{l}{ACs}&  \multicolumn{2}{|l}{Resistive heater}  & \multicolumn{2}{|l|}{Copiers}\\\cline{2-7} &CP& ZIP & CP& ZIP & CP& ZIP \\ \hline
       18 & 100 & 282 &66 & 163&59 & 151\\ \hline
       22 & 3998 & 6127 &2713 &3169 &2428 &2884 \\ \hline
        25 & 3433 & 5251 &2068 &2441 &2003 &2270\\ \hline
        33 & 327 & 1177 &214 &481 &193 &447 \\ \hline
         \end{tabular}
             \vspace{-6mm}
\label{allbuses}
\end{table}
\section{Conclusions}\label{conclusion}
In this paper, we have investigated the impact of load altering attacks (LAAs) on the voltage profile of distribution systems with ZIP loads. Our analysis highlighted the impact of the location of LAA for two different load models (CP and ZIP). {We proposed a closed-form method to compute the bus voltages with the ZIP load model which has a reduced complexity but suffers from an accuracy loss, because assumptions needed to obtain the closed-form expressions.} Our future research will focus on investigating the impact of LAAs in more complex systems that also contain distributed generation (e.g., solar panels). Additionally, we will devise strategies to mitigate the adverse effects of the LAAs.
\vspace{-1mm}
\bibliography{IEEEabrv,Refs}
\end{document}